\documentclass[compsoc,conference,a4paper,10pt,times]{IEEEtran}
\usepackage{cite}
\usepackage{amsmath,amssymb,amsfonts}
\usepackage{graphicx}
\usepackage{textcomp}
\usepackage{bmpsize}
\usepackage{xcolor}
\usepackage{lipsum}
\usepackage[colorlinks=true,urlcolor=black]{hyperref}
\def\BibTeX{{\rm B\kern-.05em{\sc i\kern-.025em b}\kern-.08em
    T\kern-.1667em\lower.7ex\hbox{E}\kern-.125emX}}

\usepackage{float}
\usepackage{makecell}
\usepackage{tikz}
\usepackage{amsmath}
\usepackage{bm}
\usepackage{enumitem}
\usepackage[table]{xcolor}
\usepackage[htt]{hyphenat}
\usepackage{lipsum}
\usepackage{afterpage}
\usepackage{tabularx}
\usepackage{mathtools}
\usepackage{multirow}
\usepackage{hhline}
\usepackage{algorithm}
\usepackage{algpseudocode}
\usepackage{url}
\usepackage{multicol}
\algnewcommand{\LeftComment}[2][\algorithmicindent]{\State \hspace{#1} \(\triangleright\) #2}

\usepackage{tablefootnote}

\begin{document}

\title{ThreadFuzzer: Fuzzing Framework for Thread Protocol}


\author{\IEEEauthorblockN{Ilja Siroš}
	\IEEEauthorblockA{\textit{COSIC, KU Leuven} \\
		Leuven, Belgium \\
		ilja.siros@esat.kuleuven.be\ } \\
    \IEEEauthorblockN{Dave Singelée}
	\IEEEauthorblockA{\textit{DistriNet, KU Leuven} \\
		Leuven, Belgium \\
		dave.singelee@kuleuven.be\ }
	\and
    \IEEEauthorblockN{Jakob Heirwegh}
    \IEEEauthorblockA{\textit{COSIC, KU Leuven} \\
    Leuven, Belgium \\
    jakob.heirwegh@esat.kuleuven.be\ } \\
	  \IEEEauthorblockN{Bart Preneel}
	\IEEEauthorblockA{\textit{COSIC, KU Leuven} \\
		Leuven, Belgium \\
		bart.preneel@esat.kuleuven.be}
}

\maketitle

\begin{abstract}
With the rapid growth of IoT, secure and efficient mesh networking has become essential. Thread has emerged as a key protocol, widely used in smart-home and commercial systems, and serving as a core transport layer in the Matter standard.
This paper presents ThreadFuzzer, the first dedicated fuzzing framework for systematically testing Thread protocol implementations. By manipulating packets at the MLE layer, ThreadFuzzer enables fuzzing of both virtual OpenThread nodes and physical Thread devices. The framework incorporates multiple fuzzing strategies, including \textit{Random} and \textit{Coverage-based} fuzzers from CovFuzz, as well as a newly introduced \textit{TLV Inserter}, designed specifically for TLV-structured MLE messages.
These strategies are evaluated on the OpenThread stack using code-coverage and vulnerability-discovery metrics. The evaluation uncovered five previously unknown vulnerabilities in the OpenThread stack, several of which were successfully reproduced on commercial devices that rely on OpenThread. Moreover, ThreadFuzzer was benchmarked against an oracle AFL++ setup using the manually extended OSS-Fuzz harness from OpenThread, demonstrating strong effectiveness. These results demonstrate the practical utility of ThreadFuzzer while highlighting challenges and future directions in the wireless protocol fuzzing research space.
\end{abstract}


\section{Introduction}
In recent years, the Internet of Things (IoT) has witnessed explosive growth, driving demand for reliable, low-power, and secure networking solutions. Among the myriad connectivity protocols, Thread has emerged as a leading IPv6-based mesh technology tailored for smart-home and commercial IoT applications, achieving over 800 Thread-certifications and more than 300 products available in the market as of Q2 2025~\cite{Thread_Q2_2025}. Its popularity truly soared with the launch of Matter in late 2022—a cross-industry, IP-based smart-home standard that defines the application-layer language for device interactions while relying on Thread’s self-healing, low-power IPv6 mesh as one of its primary transports. Moreover, in 2017, Google has released an open-source implementation of the Thread protocol called OpenThread~\cite{OpenThread}, which is actively maintained and widely promoted for use in real-world products by device manufacturers.

As Thread adoption grows across IoT devices, ensuring the security of its implementations is critical. Yet, research on analyzing Thread’s security remains surprisingly limited. In particular, there are no known academic publications that explore the use of fuzzing—one of the most widely used techniques for testing the robustness of protocol implementations—against Thread stacks.

This paper presents the first systematic effort to develop a fuzzing framework for the Thread protocol, aimed at uncovering vulnerabilities in real-world implementations. Specifically, ThreadFuzzer targets the Mesh Link Establishment (MLE) layer, a component introduced exclusively for the Thread protocol and responsible for establishing and configuring secure radio links within a Thread network, making it the most relevant layer for this work. To enable the fuzzing, the framework leverages the open-source OpenThread implementation to generate valid Thread packets. These packets are intercepted during the construction phase on the MLE layer and manipulated as part of the fuzzing workflow.

In this paper, within the broader context of fuzzing the Thread protocol, established techniques from prior network protocol fuzzing research are revisited, and new methods specifically tailored to the structure of the MLE protocol—particularly its TLV-based message format—are introduced. The proposed framework is primarily evaluated on the widely adopted OpenThread implementation using the simulation-based fuzzing of their virtual nodes. In addition, a proof-of-concept is presented for applying our fuzzing approach to physical Thread devices, reproducing several vulnerabilities discovered during the OpenThread evaluation. 

The contributions of this paper are summarized as follows:
\begin{itemize}[leftmargin=*]
    \item This work introduces ThreadFuzzer—the first fuzzing framework specifically designed for testing implementations of the Thread protocol. The framework enables the fuzzing of both the widely used OpenThread stack and commercial Thread devices by intercepting and manipulating Thread packets at the MLE layer.

    \item The framework is primarily evaluated on the OpenThread implementation, resulting in the discovery of five previously unknown vulnerabilities. Several of these were successfully reproduced on commercial devices that rely on OpenThread.
    
    \item The evaluation also includes a comparison of multiple fuzzing strategies, including \textit{Random} and \textit{Coverage-based} fuzzers from CovFuzz~\cite{CovFuzz}, as well as \textit{TLV Inserter} introduced in this work. The results offer insights into the types of crashing inputs that different fuzzers are capable of producing.

    \item The OSS-Fuzz harness provided by OpenThread was manually extended to support stateful fuzzing with the AFL++ engine, enabling its use as a strong oracle for comparison. Although this form of instrumentation is generally impractical—requiring modifications to the target code and making it unusable for physical devices—the benchmark highlights ThreadFuzzer’s effectiveness: even in a fully \textit{black-box} mode, it discovered five of the six vulnerabilities detected by the oracle, despite the inherent limitations of interception-based approaches commonly used in network protocol fuzzers.
    
\end{itemize}

\textbf{Responsible Disclosure and Open Sources}. As part of the responsible disclosure process, Google was notified of the findings via their Bug Bounty program~\cite{googleBugHunters} and promptly fixed the identified vulnerabilities. To facilitate further research and investigations in the field, the code of the fuzzing framework is publicly available at \url{https://github.com/KULeuven-COSIC/ThreadFuzzer}.

\section{Background}
\subsection{Matter}
Matter is an open standard for smart-home technology to connect IoT devices to any Matter-certified ecosystem using a single protocol. Matter comes from the Connectivity Standards Alliance, an organization of hundreds of companies creating products for the smart home.
Building with Matter offers several advantages: it provides lower latency and higher reliability through IP-based local connectivity, reduces development costs by enabling cross-ecosystem compatibility with a single implementation, ensures a consistent setup experience across devices, and simplifies development by eliminating the need for account linking. Existing integrations can also be upgraded with Matter support using tools provided by Google~\cite{matter-overview}.

At the core of Matter's design is the cluster model, which structures device functionality into a standardized set of attributes, commands, and events. A cluster represents a specific feature or capability—such as a light switch, temperature sensor, or door lock—and defines how that capability is described and controlled across all Matter-compliant devices. Each cluster consists of server and client roles, with well-defined behaviors to ensure consistent interactions regardless of manufacturer.

Matter relies on two main connectivity technologies: Thread and Wi-Fi. This has been a major driver of Thread’s growing adoption, and with that growth comes the need for tools to evaluate Thread’s security—precisely what this work aims to address.

\subsection{Thread}
\subsubsection{Overview}
Thread is an IPv6-based wireless mesh networking protocol built on IEEE 802.15.4-2006, designed for low-power IoT devices within Wireless Personal Area Networks (WPANs). Unlike ZigBee, Z-Wave, or Bluetooth LE, Thread operates independently and offers key features such as simple setup, strong security through device authentication and encrypted communication, self-healing and interference-resistant reliability, energy efficiency enabling years of battery operation, and scalability to support networks with hundreds of devices~\cite{Thread_Overview}.

\subsubsection{Device roles and types}
In a Thread network, devices fulfill one of two forwarding roles - Routers, which keep their transceivers enabled at all times, forward packets for other nodes, and provide secure commissioning services; and End Devices, which communicate solely with a single parent Router and may disable their radios to conserve power. This parent–child relationship ensures that End Devices attach to exactly one Router, while Routers form the resilient mesh backbone.

Beyond these roles, Thread defines two device types based on resource capabilities. Full Thread Devices (FTDs) always maintain IPv6 address mappings, subscribe to the all-routers multicast address, and include Routers, Router-Eligible End Devices (REEDs, which can auto-promote to Routers), and Full End Devices (FEDs, which cannot). Minimal Thread Devices (MTDs) omit multicast subscriptions and forward all traffic to their parent; they appear as Minimal End Devices (MEDs, radio always on) or Sleepy End Devices (SEDs, which wake periodically to poll for messages). FTDs can switch between Router and End Device roles (e.g., a REED upgrading when it’s the only node reachable), whereas MTDs are confined to the End Device role.

Thread networks are further managed by a dynamically elected \textit{Leader}—one per partition—responsible for aggregating and distributing network-wide configuration, and by one or more \textit{Border Routers}, which bridge the Thread mesh to external IP networks. Large deployments may split into multiple partitions (each with its own Leader and Router assignments) when radio connectivity is interrupted; these partitions automatically merge upon re-establishing connectivity. To optimize performance and reliability, a partition supports up to 32 Routers and up to 511 End Devices per Router, with automatic promotion and demotion mechanisms to maintain these bounds~\cite{Thread_Roles}.

\subsubsection{Mesh Link Establishment layer}
One of the core layers in Thread is a Mesh Link Establishment (MLE) layer, which was introduced specifically in the Thread protocol specification. It is responsible for network formation and maintenance.

Its central mechanism is the \textit{MLE attach procedure}, where a new device begins the attachment process by sending a \textit{Parent Request} to nearby Routers, selects a parent based on \textit{Parent Responses}, and follows up with a \textit{Child ID Request}. If accepted, the Router replies with a \textit{Child ID Response} and propagates the child’s address within the mesh. After attachment, the device and its parent exchange \textit{Child Update Request/Response} and \textit{Data Request/Response} messages to maintain synchronization and support data flow. All these message types are generated at the MLE layer.

After joining a Thread network, a child device may promote itself to a Router. To initiate this, it sends an \textit{Address Solicit} message to the Leader, requesting a Router ID. If accepted, the Leader replies with an \textit{Address Solicit Response} containing the assigned Router ID. The device then upgrades its role to Router and broadcasts \textit{Link Requests} to neighboring Routers, which reply with either a \textit{Link Response} or a \textit{Link Accept and Request} to establish bi-directional links.

MLE packets consist of a security header, a one-byte message type, and a sequence of Type-Length-Value (TLV) records. Each TLV comprises a type, a length, and a data field—where the data may itself contain nested TLVs. All of these fields can be targeted during fuzzing.

\subsubsection{OpenThread}
At the time of writing, the primary open-source implementation of the Thread protocol is OpenThread~\cite{OpenThread}, released by Google. As a Thread-certified stack, it supports all core components of the protocol, including IPv6, 6LoWPAN, IEEE 802.15.4 MAC security, MLE, mesh routing, device roles, and Border Router functionality. OpenThread is designed for portability, featuring a minimal abstraction layer and low memory footprint, making it suitable for both system-on-chip and network co-processor architectures. Originally developed for Google Nest products, it is now used in over 100 commercial devices~\cite{OpenThread}. 

To support local testing, OpenThread provides the OpenThread Network Simulator (OTNS)~\cite{OTNS}, which allows creating virtual nodes and accelerates communication between them through virtual time—an ability that is particularly useful for fuzzing.
OpenThread also uses OSS-Fuzz~\cite{OSS_Fuzz} to fuzz-test its implementation. The limitations of this approach, and how it compares to ThreadFuzzer, are discussed in Section~\ref{subsec:ComparisonWithOSSFuzz}.



\subsection{Fuzzing}\label{sec:Background_Fuzzing}

Fuzzing, first introduced by Miller et al.~\cite{Miller_Fuzzing}, is a fundamental technique in security testing that exposes software to random or malformed inputs to uncover crashes and unexpected behavior often missed by traditional tests.

A foundational tool in this space is American Fuzzy Lop (AFL)~\cite{AFL}, a coverage-guided fuzzer that uses instrumentation to track executed code paths and address sanitization to detect memory errors at runtime. AFL’s methodology has shaped the design of many modern fuzzers, including its successor AFL++~\cite{aflplusplus},

Fuzzing wireless protocols, however, introduces added complexity. These protocols rely on strict message formats, integrity checks, encryption, and stateful behavior—making input mutation and valid packet timing far more challenging. Since inputs must reach the device in specific protocol states, fuzzers must account for timing and sequencing.

In general, there are two distinct ways to approach the fuzz testing of wireless protocols in physical devices: emulation-based fuzzing and over-the-air fuzzing. Emulation is much more efficient than over-the-air testing: iterations run faster, and information like coverage data is available directly from the emulator. However, creating a working emulation of a device is difficult and requires a lot of manual effort.

Besides being much slower, over-the-air fuzzing faces an important limitation: restricted observability and the absence of instrumentation make feedback collection difficult, reducing fuzzing efficiency. Certain classes of bugs may become practically impossible to detect. Memory-corruption issues, for instance, can be especially hard to observe because their effects may be subtle, delayed, or produce no clear crash signature~\cite{What_You_Corrupt_Is_Not_What_You_Crash}.

The main advantage of over-the-air fuzzing, however, is its robustness across targets—the same framework can be applied to devices from different vendors and targets using different protocol implementations. This portability was the primary motivation for developing ThreadFuzzer as an over-the-air fuzzing framework.


\section{Related work}
In recent years, significant progress has been made in network protocol fuzzing. Frameworks such as AFLNet~\cite{AFLNet}, StateAFL~\cite{StateAFL}, ChatAFL~\cite{chatafl}, LibAFLStar~\cite{LibAFLStar} and SGFuzz~\cite{SGFuzz} extend AFL++~\cite{aflplusplus} or LibFuzzer~\cite{libfuzzer} to support stateful network-driven and event-based targets. As a result, they rely on coverage-feedback mechanisms and operate in a \textit{grey-box} or even \textit{white-box} fashion.

Although these tools significantly enhance network protocol fuzzing, they inherently rely on software instrumentation and often require modifications to the target source code (e.g., removing process forks, disabling encryption and integrity checks). Consequently, these approaches are impractical for \textit{black-box} fuzzing of physical devices or in scenarios where the target software cannot be instrumented or modified.

Parallel to these efforts, researchers have explored wireless protocol fuzzing, targeting diverse technologies including Bluetooth~\cite{SweynTooth, Frankenstein, BrakTooth, Stateful_BL_Fuzz, L2Fuzz, BTFuzzer}, Wi-Fi~\cite{Greyhound, Owfuzz, Wifi_test_and_fuzz_framework}, 4G/5G~\cite{LTEFuzz, BERSEKER, CovFuzz, U-Fuzz} and Zigbee~\cite{U-Fuzz, Z-Fuzzer, ZigBee_Fuzz_Constraint, LLMIF}. These works utilize either the emulation-based fuzzing or the over-the-air one to identify vulnerabilities in the physical device implementations.

Despite the growing adoption of the Thread protocol in IoT ecosystems, its security research remains sparse. One of the few prior works~\cite{threadSec} repurposed Zigbee analysis tools on development boards running OpenThread to dissect Thread traffic and evaluate vulnerabilities to energy-depletion and online password-guessing attacks. Subsequent research~\cite{threadSymbolic}, performed symbolic verification of Thread’s Mesh Commissioning Protocol. Nevertheless, to date, no one has applied fuzz-testing techniques to the Thread protocol, leaving a significant gap in its security analysis, which this work aims to bridge.

\section{Framework architecture} \label{sec:architecture}
The fuzzing framework combines elements of both generation- and mutation-based fuzzing by leveraging an open-source Thread stack to generate protocol-compliant packets. During this process, packets are intercepted at the MLE layer and selectively mutated. This approach directly addresses the challenges of fuzzing stateful network protocols discussed in Section~\ref{sec:Background_Fuzzing}. By restricting mutations to a single protocol layer, the framework ensures that the underlying layers remain valid, allowing the mutated packets to pass encryption/integrity checks and reach the target’s parser for the manipulated layer. Furthermore, the use of a protocol stack ensures that packets are sent in the correct sequence and at appropriate times, significantly improving fuzzing efficiency by maintaining valid protocol state throughout the session. Despite its limitations, which are discussed in Section~\ref{subsec:benign_packet_limitation}, the interception-based strategy has proven effective in several prior fuzzers targeting wireless protocols~\cite{CovFuzz, U-Fuzz, Greyhound, Greyhound_for_4G_5G}.

ThreadFuzzer builds upon the CovFuzz~\cite{CovFuzz} architecture—originally developed for 4G and 5G protocol fuzzing—adapting its design for fuzzing the Thread protocol, including reuse of certain fuzzing components described in Section~\ref{subsec:fuzzer}.

\begin{figure*}
    \centering
    \includegraphics[width=\textwidth]{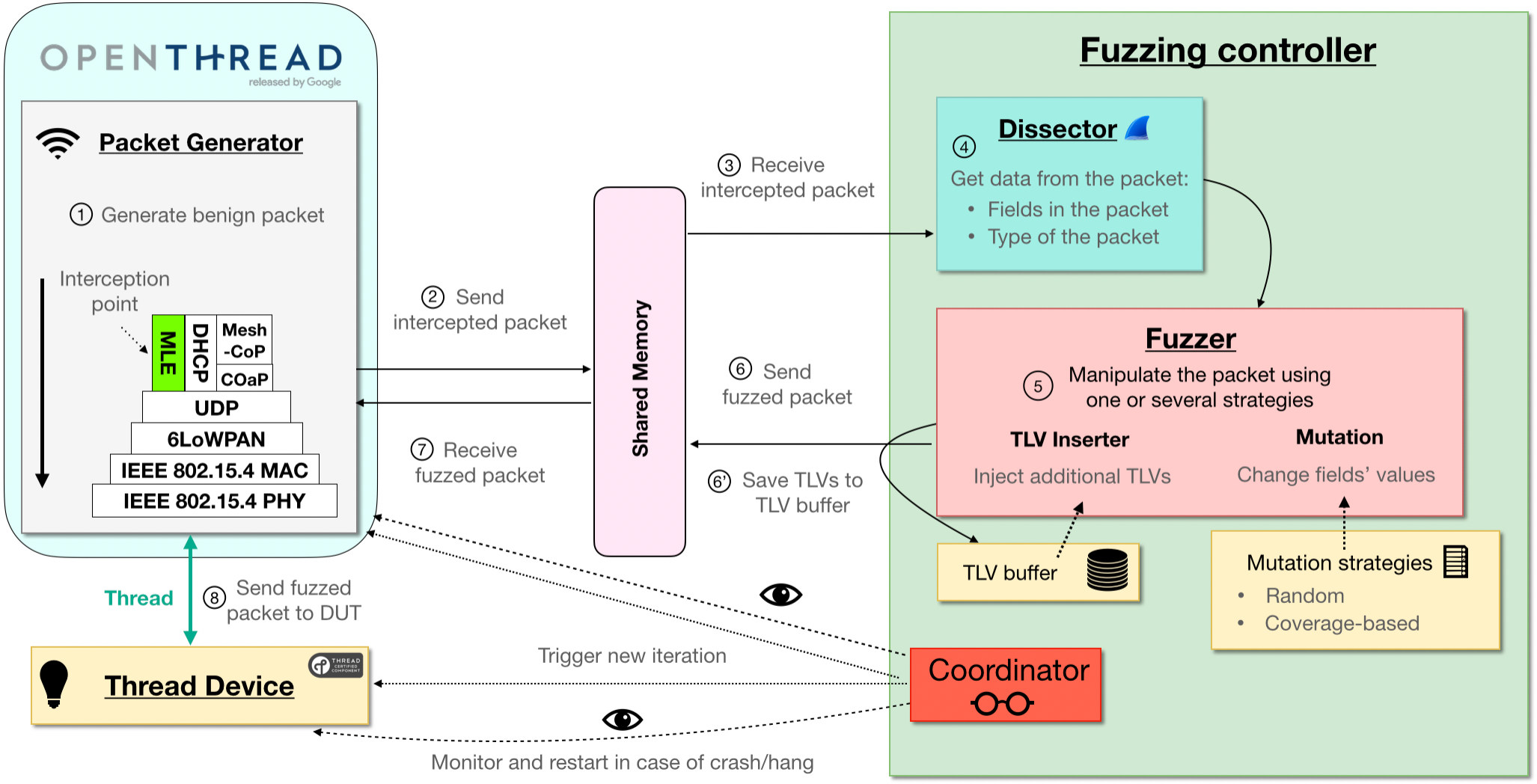}
    \caption{The illustration of the ThreadFuzzer. The steps shown in the figure outline the fuzzing flow for one packet. The figure is adapted from~\cite{CovFuzz}.}
    \label{fig:framework_architecture}
\end{figure*}

ThreadFuzzer consists of three primary components: \textit{Packet Generator} (\textit{PG}), \textit{device-under-test} (\textit{DUT}), and \textit{Fuzzing Controller}. Below, a detailed description of the functionality and responsibilities of each component is provided with a clear distinction between simulation-based fuzzing of OpenThread nodes and fuzzing of commercial Thread devices.

\subsection{Packet Generator} \label{subsec:packetgenerator}
In ThreadFuzzer, the \textit{Packet Generator} (\textit{PG}) is responsible for producing benign Thread packets. To enable fuzzing, the \textit{PG} code is instrumented by inserting software hooks at the MLE layer. These hooks intercept packets during their construction phase, forward them to \textit{Fuzzing Controller} via a shared memory interface, and receive the modified packets for eventual transmission to \textit{DUT}.

The implementation of \textit{PG} in ThreadFuzzer varies depending on the type of \textit{DUT}. Empirical evaluation showed that the OpenThread Border Router (OTBR), implemented in \texttt{ot-br-posix}~\cite{Ot-br-posix} and built on top of the main OpenThread library, provides the broadest support for message types and optional TLV fields across different packet formats. As a result, OTBR is used as \textit{PG} when fuzzing physical devices.

However, OTBR lacks a critical feature required for efficient fuzzing of virtual nodes: support for virtual discrete-time simulation, which significantly accelerates inter-node communication. This feature is available only in OpenThread's implementations of the  MTD (OT-MTD) and FTD (OT-FTD). Therefore, to maximize fuzzing performance in simulated environments, ThreadFuzzer uses OT-FTD as \textit{PG} when targeting virtual Thread nodes.

To approximate the behavior of OTBR, OT-FTD is compiled with specific flags used in the OpenThread Network Simulator (OTNS)~\cite{OTNS}. Experimental results confirm that the modified OT-FTD emits a wide range of message types and optional fields—comparable to those generated by OTBR—even when the associated services are not explicitly invoked by any higher-level application.

Additionally, to automate control of virtual nodes during fuzzing, ThreadFuzzer employs UNIX named pipes to issue commands such as \texttt{factoryreset} and \texttt{restart}. This approach eliminates the need for manual interaction via the command-line interface, thereby improving testing efficiency.

\subsection{DUT}\label{subsec:DUT}
\textit{DUT} is the fuzzing target.
As mentioned before, ThreadFuzzer supports fuzzing both virtual and physical \textit{DUT}s.

Fuzzing virtual nodes offers several advantages that significantly enhance efficiency. First, as mentioned in Section~\ref{subsec:packetgenerator}, the OpenThread library includes a speedup mechanism for virtual nodes that allows to drastically improve throughput. Second, since each node runs as a separate application, monitoring their status is straightforward—it is enough to check whether the corresponding process is still active. Finally, using an open-source \textit{DUT} makes it possible to instrument the software with address and coverage sanitizers. This enables the easy detection of memory corruptions and the collection of code coverage data, which can be used both to guide the fuzzing process and serve as an additional performance metric.

On the other hand, fuzzing commercial devices introduces additional challenges due to the absence of built-in instrumentation and limited observability. In particular, reliably detecting crashes during execution requires an external indicator. This work leverages on the reboot-count attribute of Matter’s \textit{General Diagnostics cluster} as a proxy for unexpected reboots.

\begin{algorithm}[H]
\caption{Pseudocode of a single fuzzing epoch for a physical Thread device.}
\label{alg:realDeviceFuzzingProcedure}
\begin{algorithmic}[1]
\Procedure{RunFuzzingEpoch}{$N$}
\State \Call{PerformHardReset}{$ $}
\State $C_0 \gets$ \Call{ReadRebootCount}{$ $}
\State \Call{PerformSoftReset}{$ $}
\For{$i = 1$ to $N$}
\State \Call{RunFuzzingIteration}{$ $}
\State \Call{PerformSoftReset}{$ $}
\EndFor
\State \Call{RunCleanAttach}{$ $}
\State $C_f \gets$ \Call{ReadRebootCount}{$ $}
\State \Return $C_f > C_0 + N + 1$
\EndProcedure
\end{algorithmic}
\end{algorithm}

The overall workflow is outlined in Algorithm~\ref{alg:realDeviceFuzzingProcedure}. Fuzzing proceeds in epochs. At the start of each epoch, a hard reset is performed. This reboots the device, applies a factory reset, and clears the reboot-count. The initial reboot-count \(C_0\) is then read from the device. The epoch proceeds with \(N\) fuzzing iterations, each consisting of an MLE Attach procedure in which mutated packets are injected. Every iteration ends with a soft reset—which simply power-cycles the device—to trigger the next attach attempt. A soft reset is also performed before the first iteration for the same reason. After completing the \(N\) fuzzing iterations, one clean (non-fuzzed) iteration is executed to allow \textit{DUT} to rejoin the Thread network, after which the final reboot-count \(C_f\) is recorded.

The framework itself triggers exactly \(N + 1\) soft resets per epoch. If the observed value satisfies \(C_f > C_0 + N + 1\), an unexpected reboot occurred, indicating that at least one fuzzed input caused a crash. To identify the responsible input, the \(N\) iterations can be replayed individually. A smaller choice of \(N\) narrows the search space when replaying crashing seeds but increases the proportion of hard resets, reducing overall fuzzing throughput.

Note the distinction between a soft and a hard reset. A soft reset may allow some state to leak between iterations, whereas a hard reset guarantees a complete device reset with no leakage. Consequently, the epoch size \(N\) represents a trade-off between higher overhead (\(N = 1\)) and reduced stability (\(N > 1\)).

Device pairing over Matter and reboot‐count inspection are both performed using \texttt{chip-tool}, the open‐source Matter controller included in the Matter reference implementation~\cite{Matter}.

\subsection{Fuzzing Controller}

\textit{Fuzzing Controller} manages the core fuzzing operations and orchestrates the entire fuzzing workflow. Its responsibilities include mutating intercepted packets, monitoring the behavior of both \textit{DUT} and \textit{PG}, and handling recovery actions such as restarting components when failures occur. It also manages fuzzing iterations, logs execution details, and collects coverage data from \textit{DUT} and \textit{PG} when available. It consists of the following primary sub-components:

\subsubsection{Dissector}
\textit{Dissector} handles packet parsing by identifying the packet type and extracting its fields. This is implemented using a custom library built on top of Wireshark~\cite{Wireshark}, which provides robust support for dissecting network protocols. The same library has been used in other network protocol fuzzing frameworks~\cite{CovFuzz, U-Fuzz}.
Once the packet type and the corresponding fields are identified, the packet can be more intelligently fuzzed by the \textit{Fuzzer}.

\subsubsection{Fuzzer} \label{subsec:fuzzer}
\textit{Fuzzer} is an abstract class that defines the interface for manipulating a packet according to a specific fuzzing strategy. In ThreadFuzzer, \textit{Random} and \textit{Coverage-based} fuzzers are borrowed from CovFuzz~\cite{CovFuzz} and \textit{TLV Inserter} is introduced as a new fuzzer. It is a specialized fuzzer that is specifically targeted towards TLV-encoded structures. The following paragraphs provide a detailed description of each fuzzer.

\textit{\textbf{Random fuzzer}} introduced in CovFuzz, alters the values of specific fields within the intercepted packet.
Every field in the packet gets assigned a mutation probability, calculated as follows:
\[
    p_f = \frac{k}{|F_P|}
\]
Where:
\begin{itemize}[leftmargin=*]
    \item $p_f$ is a probability of mutating a given field
    \item $|F_P|$ is the total number of fields in the packet
    \item $k$ is a hyperparameter representing the average number of mutated fields per packet
\end{itemize}
This approach ensures that the probability is evenly distributed across all fields in a packet, and the average number of mutated fields per packet remains the same among all the intercepted packets. The hyperparameter \(k\) should be selected empirically, depending on the sensitivity of the fuzzed protocol to mutated inputs.

\textit{\textbf{Coverage-based fuzzer}}, originally introduced in CovFuzz~\cite{CovFuzz}, extends \textit{Random fuzzer} by incorporating coverage feedback to guide mutations. It dynamically adjusts the field mutation probabilities \(p_f\) to maximize code coverage of \textit{DUT}.
The fuzzer is parameterized by two hyperparameters: \(k\) and \(\beta\). The parameter \(k\) is identical to that used in \textit{Random fuzzer} and is used to compute the initial mutation probabilities for each field. The parameter \(\beta\) controls the adaptation rate of these probabilities during fuzzing. Higher values of \(\beta\) increase the probability of mutating fields that led to new coverage and reduce the penalty when coverage remains unchanged. In contrast, lower values of \(\beta\) result in more conservative updates—yielding smaller gains when new coverage is achieved and sharper reductions when no progress is observed.
CovFuzz implements both \textit{Coverage-based Grey-box} and \textit{Coverage-based Black-box} variants of this approach. The\textit{Grey-box} version uses direct coverage feedback from \textit{DUT} to inform input selection, while the \textit{Black-box} version approximates \textit{DUT} coverage by using feedback from \textit{PG}. The \textit{Black-box} approach is particularly useful when direct coverage data from \textit{DUT} is unavailable—for instance, when fuzzing commercial devices. CovFuzz demonstrated the effectiveness of this technique in the context of 4G protocol fuzzing. For additional details, see CovFuzz~\cite{CovFuzz}.

\textit{\textbf{TLV Inserter}} is specifically designed to enhance fuzzing of TLV-structured packets. It operates by inserting previously extracted TLVs into intercepted packets at valid boundaries (i.e., between existing TLVs or as a sub-TLV in the existing TLV), thereby preserving syntactic correctness. To maintain internal consistency, \textit{TLV Inserter} includes a mechanism for updating the length fields of any affected “parent” TLVs, controlled by a probability parameter~$\gamma$. When $\gamma = 1$, all relevant length fields are recalculated to reflect the modified structure, ensuring structural validity of the packet. Conversely,  when $\gamma = 0$, no length fields are updated. Setting $\gamma$ to a value below 1 intentionally introduces occasional inconsistencies between length and value fields. Although such inconsistencies may cause the Thread parser to reject the packet, they can also expose parsing bugs that arise when developers fail to handle malformed inputs properly.

To ensure that the inserted TLVs are themselves subject to mutation, \textit{TLV Inserter} is used in conjunction with \textit{Random fuzzer}. This can be easily done as the framework supports chaining multiple fuzzers in sequence. The composition of fuzzer modules can be specified through a configuration file, allowing for flexible experimentation.

\subsubsection{Coordinator}
The \textit{Coordinator} manages the lifecycle of fuzz\-ing iterations by initiating and terminating test cycles. It also monitors the status of both \textit{PG} and \textit{DUT}. In the event of anomalies or crashes, the \textit{Coordinator} ensures that the fuzzing process is promptly restarted.
ThreadFuzzer implements a \textit{Timeout-based Coordinator}, which assigns a fixed time budget to each fuzzing iteration. Once the timeout expires, the iteration is forcibly terminated. The timeout value is selected to strike a balance between two competing goals: maximizing the number of distinct messages observed (i.e., increasing protocol coverage), and minimizing idle periods without message activity (i.e., reducing ineffective waiting time). The fuzzing flow from the Coordinator's perspective is shown in Algorithm~\ref{alg:fuzzingFlow}.

\begin{algorithm}
    \caption{Fuzzing flow in the \textit{Timeout-based Coordinator}.}\label{alg:fuzzingFlow}
    \begin{algorithmic}[1]
        \State \textbf{Input:} 
        \State \hspace{1em} $max\_i$ \Comment{Maximum number of iterations}
        \State \hspace{1em} $it\_len$ \Comment{Fuzzing iteration length}
        \State
        \For {$i$ in $1,2,\dots, max\_i$}
            \LeftComment[0\dimexpr\algorithmicindent]{Activate Thread in \textit{PG} and \textit{DUT}}
            \State \Call{activateThread}{$ $}
            \State \Call{startTimer}{$it\_len$}
            \LeftComment[0\dimexpr\algorithmicindent]{While timer is running}
            \While {not \Call{timerIsOver}{$ $}}
            \LeftComment[0\dimexpr\algorithmicindent]{Receive the packet via shared memory}
                \State $packet \gets \ $\Call{shm.recv}{$ $} 
            \LeftComment[0\dimexpr\algorithmicindent]{Dissect the packet}
                \State \Call{dissect}{$packet$}
            \LeftComment[0\dimexpr\algorithmicindent]{Fuzz the packet}
                \For{fuzzer in fuzzers}
                \State fuzzer.\Call{fuzz}
                {$packet$}
                \EndFor
            \LeftComment[0\dimexpr\algorithmicindent]{Send the packet back to interception point}
                \State \Call{shm.send}{$packet$}
            \EndWhile
            \LeftComment[0\dimexpr\algorithmicindent]{Check if PG or DUT has crashed}
            \If {not DUT.\Call{isAlive}{$ $}} \LeftComment[0\dimexpr\algorithmicindent]{Log the crash}
                \State logger.\Call {log}{``DUT crashed!''}
                \LeftComment[0\dimexpr\algorithmicindent]{Restart the \textit{DUT}}
                \State DUT.\Call{restart}{$ $}
            \EndIf
            \LeftComment[0\dimexpr\algorithmicindent]{Get coverage information (if possible)}
            \State \Call {getCoverageFeedback}{$ $}
            \LeftComment[0\dimexpr\algorithmicindent]{Call factory reset on nodes}
            \State \Call{factoryReset}{$ $}
        \EndFor
        
    \end{algorithmic}
\end{algorithm}


\section{Evaluation}

In evaluating ThreadFuzzer, the following research questions are addressed:

\begin{itemize}
\item[\textbf{RQ1}] Can ThreadFuzzer identify vulnerabilities in the Thread protocol implementation?
\item[\textbf{RQ2}] Can vulnerabilities discovered in OpenThread be reproduced on commercial Thread devices?
\item[\textbf{RQ3}] How efficient are the fuzzers introduced in CovFuzz when applied to the Thread protocol?
\item[\textbf{RQ4}] How efficient is the TLV Inserter which is introduced in this paper?
\item[\textbf{RQ5}] How efficient is ThreadFuzzer framework in fuzzing the Thread protocol implementation?
\end{itemize}

\textbf{RQ1} is addressed in Section~\ref{subsec:ThreadEval}. \textbf{RQ3} and \textbf{RQ4} are examined in Section~\ref{subsec:CompareFuzzers}. \textbf{RQ5} is analyzed in Section~\ref{subsec:ComparisonWithOSSFuzz}, and \textbf{RQ2} is covered in Section~\ref{subsec:CommercialFuzz}.

\subsection{Experimental setup}
All experiments were conducted on an Intel Core i7 HP EliteBook 850 G8 with 16\,GB of RAM and 8 CPU cores, running Ubuntu 22.04. The OpenThread version used for all experiments was \texttt{thread-reference-20230706}, the latest official release available at the time of the research.
Experiments involving commercial \textit{DUT}s additionally employed the OpenThread Border Router~\cite{Ot-br-posix} (version \texttt{thread-reference-20230710}) in conjunction with the nRF52840 Development Kit~\cite{NRF52840DK}. The development kit includes a Thread-capable chip that serves as a radio co-processor, enabling communication with \textit{DUT} over the Thread protocol.

\subsection{ThreadFuzzer Evaluation on OpenThread Virtual Targets}\label{subsec:ThreadEval}
ThreadFuzzer is evaluated on two OpenThread virtual targets: OT-FTD and OT-MTD. Each fuzzer's performance is first assessed on the OT-FTD target using two primary metrics: code coverage and the number of unique vulnerabilities discovered. Based on these results, the most effective fuzzers are subsequently applied to OT-MTD.
All experiments with identical settings were repeated at least five times, each consisting of 10,000 fuzzing iterations, to ensure statistical significance.

\subsubsection{Overview of the  findings}
In total, five vulnerabilities were identified during fuzzing: four reachable assertion failures and one stack buffer overflow. Of these, four are reproducible on the OT-FTD target and three on the OT-MTD target. A summary of these vulnerabilities is provided in Table~\ref{tab:ot_vulns}, which reports the fuzzed message, type of manipulation, target build (FTD or MTD), and the crash reason. Each crash is assigned a distinct identifier (\textbf{C1}–\textbf{C5}) for ease of reference. The causes for the crashes are analyzed in detail in Section~\ref{subsec:crashAnalysis}.

Overall, the preceding analysis supports the conclusion that, in response to \textbf{RQ1}, ThreadFuzzer can identify vulnerabilities in the Thread protocol implementation.

\begin{table*}[htb]
\caption{Vulnerabilities found during the fuzzing}\label{tab:ot_vulns}
\small
\begin{tabular}{|c|c|c|c|c|}
\hline
\textbf{Name} & \textbf{Message} & \textbf{Packet manipulation} & \textbf{DUT} & \textbf{Reason} \\
\hline
\textbf{C1} & Child ID Response & Set \textit{Prefix Length} in Prefix TLV to \textit{255} &  \makecell{OT-FTD \\ OT-MTD} & Reachable Assertion \\
\hline
\textbf{C2} & Child ID Response & \makecell{Set \textit{Length} in Server TLV to \textit{1} \\ Set \textit{Address16} equal to \textit{Server16}} & OT-FTD & Stack Buffer Overflow \\
\hline
\textbf{C3} & \makecell{Data Response or\\Child ID Response} & Set \textit{Length} in Network Data TLV to \textit{255} & \makecell{OT-FTD \\ OT-MTD} & Reachable Assertion \\
\hline
\textbf{C4} & Child Update Response & Set \textit{Timeout} in Timeout TLV to \textit{4294967295} & OT-MTD & Reachable Assertion \\
\hline
\textbf{C5} & Advertisement & Set \textit{Leader ID} in Leader Data TLV to \textit{255} & OT-FTD & Reachable Assertion \\
\hline

\end{tabular}
\end{table*}

\subsubsection{Comparative Evaluation of Fuzzer Variants} \label{subsec:CompareFuzzers}
As outlined in Section~\ref{subsec:fuzzer}, ThreadFuzzer integrates multiple fuzzing strategies. To address \textbf{RQ3} and \textbf{RQ4}, each fuzzer is evaluated on the OT-FTD target. The evaluation includes two fuzzers adapted from CovFuzz, as well as a new fuzzer introduced in this work:

\begin{itemize}[leftmargin=*]
  \item \textbf{\textit{Random fuzzer}}: Serves as the baseline for comparison. Preliminary experiments identified \(k = 2\) as an effective value for the \(k\)-hyperparameter, which is fixed for all subsequent evaluations.
  
  \item \textbf{\textit{Coverage-based fuzzer}}: Builds on \textit{Random fuzzer} by incorporating coverage feedback. The same \(k = 2\) setting is used, and multiple configurations of the weighting parameter \(\beta\) are explored, following guidance from the CovFuzz study. Both \textit{Grey-box} and \textit{Black-box} variants are evaluated.
  
  \item \textbf{\textit{TLV Inserter}}: Introduced in this work, this fuzzer targets the TLV-based structure of Thread packets. Experiments are conducted with varying values of the \(\gamma\) parameter, and the results are compared against \textit{Random fuzzer} baseline.
\end{itemize}

\begin{figure}
    \centering
    \includegraphics[width=\columnwidth]{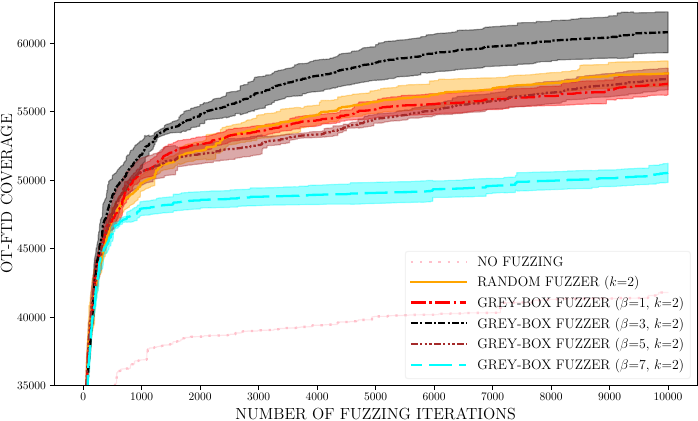}
    \caption{Comparison of \textit{Random} and \textit{Coverage-based Grey-box fuzzer}s from CovFuzz.}
    \label{fig:Grey-box_fuzzing}
\end{figure}

\begin{figure}
    \centering
    \includegraphics[width=\columnwidth]{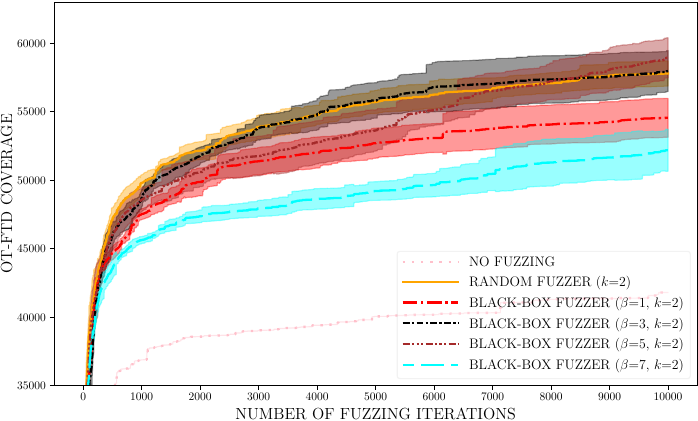}
    \caption{Comparison of \textit{Random} and \textit{Coverage-based Black-box fuzzer}s from CovFuzz.}
    \label{fig:Black-box_fuzzing}
\end{figure}

\begin{figure}
    \centering
    \includegraphics[width=\columnwidth]{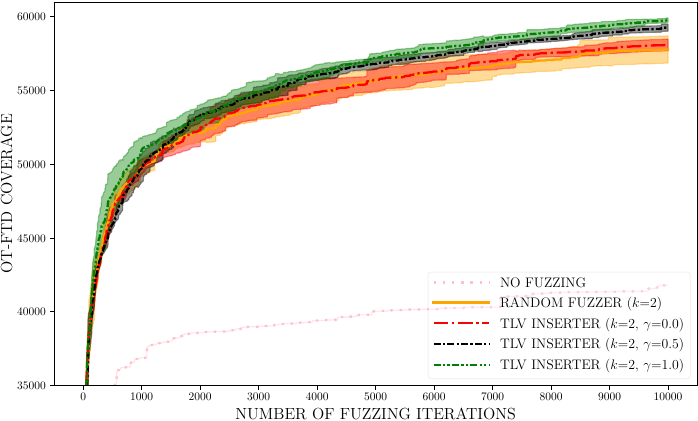}
    \caption{Comparison of \textit{Random fuzzer} and \textit{TLV inserter}.}
    \label{fig:tlv_inserter}
\end{figure}

\begin{figure}
    \centering
    \includegraphics[width=\columnwidth]{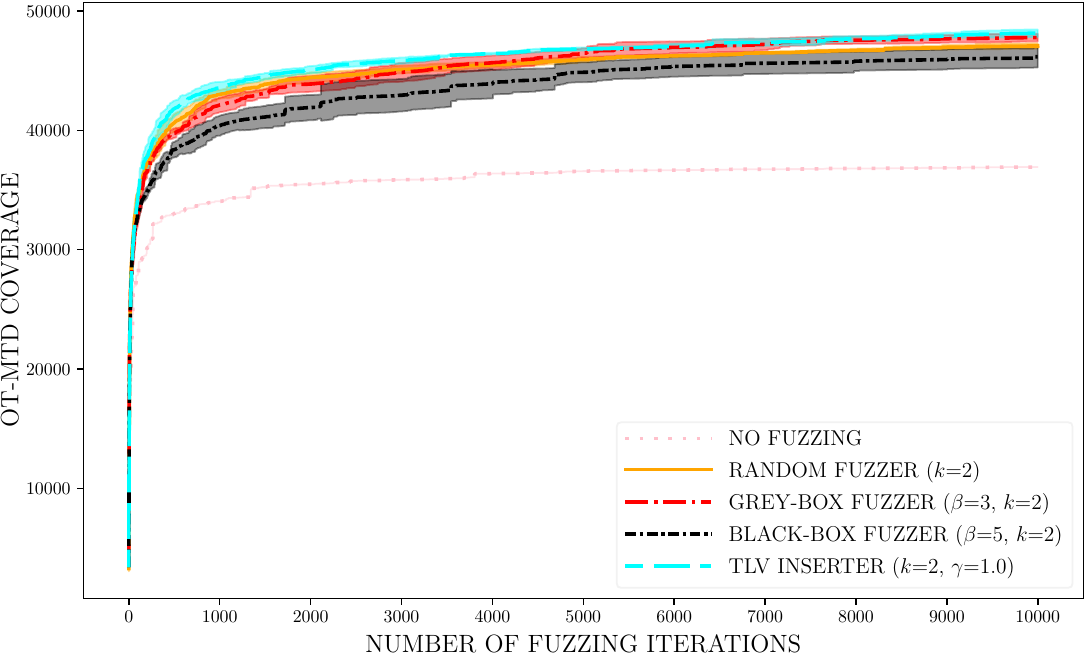}
    \caption{Comparison of fuzzers on the OT-MTD target.}
    \label{fig:mtd_fuzzing}
\end{figure}

Results for all fuzzers are first presented, followed by a comparative analysis. For the OT-FTD target, Figures~\ref{fig:Grey-box_fuzzing} and~\ref{fig:Black-box_fuzzing} show the code coverage achieved by \textit{Coverage-based Grey-Box} and \textit{Coverage-based Black-box} fuzzers, respectively. Figure~\ref{fig:tlv_inserter} compares the performance of \textit{TLV Inserter} across different values of the \(\gamma\) parameter. Subsequently, Figure~\ref{fig:mtd_fuzzing} presents the results of the best-performing fuzzers when applied to the OT-MTD target. 
For reference, all figures include results from \textit{Random fuzzer} baseline and a ``No Fuzzing'' configuration, which serves as a 0-coverage reference point.
In addition, Table~\ref{tab:CrashPerformance} summarizes each fuzzer’s performance, including the mean number of iterations (with standard deviation) required to trigger each crash, and the average execution time per iteration.

\textbf{Analysis of fuzzing OT-FTD.} While the \textit{Random fuzzer} shows a solid performance, the best \textit{Coverage-based Grey-box fuzzer} configuration (\(k=2\), \(\beta=3\)) achieves approximately an 18.8\% improvement in code coverage over \textit{Random fuzzer}\footnotemark[1]. Similarly, the most effective \textit{Coverage-based Black-box fuzzer configuration (\(k=2\), \(\beta=5\))} yields a smaller improvement of 7.2\% over the \textit{Random fuzzer}. These findings are consistent with the results reported in CovFuzz, which demonstrated that \textit{Coverage-based Grey-box} fuzzing performs the best, while \textit{Coverage-based Black-box fuzzer} still slightly surpasses \textit{Random fuzzer}. Notably, the choice of \(\beta\) has a significant impact on fuzzing efficacy, again corroborating prior observations from CovFuzz.

\footnotetext[1]{Code coverage increase is measured relative to the ``No Fuzzing" line, which is treated as having 0 coverage.}

In terms of vulnerability discovery, the \textit{Random fuzzer} detects \textbf{C1}, \textbf{C2} and \textbf{C5}. However, both best \textit{Coverage-based Grey-box} and \textit{Coverage-based Black-box fuzzer}s identified one additional vulnerability (\textbf{C3}) compared to \textit{Random fuzzer} baseline, even though \textbf{C3} was not detected in every fuzzing run.
Notably, even a fuzzer with similar coverage to \textit{the Random fuzzer} (ex. \textit{Grey-box fuzzer} with ($k=2, \beta=5$)) was able to detect \textbf{C3}, illustrating that while coverage is generally correlated with the number of bugs found, higher coverage does not always result in more vulnerabilities discovered.

For \textit{TLV Inserter}, all tested configurations outperformed \textit{Random fuzzer} to some extent. The best results were achieved with \(\gamma = 1.0\), which led to a about 12.5\% improvement in coverage. Similarly to \textit{Coverage-based fuzzer}s, \textit{TLV Inserter} managed to uncover \textbf{C3}, but the detection of it was not very reliable.

\textbf{Analysis of fuzzing OT-MTD.}
For the OT-MTD evaluation, only the best-performing fuzzers identified in the OT-FTD experiments were selected. The corresponding coverage results are presented in Figure~\ref{fig:mtd_fuzzing}. Among them, \textit{TLV Inserter} demonstrated the highest coverage, improving upon \textit{Random fuzzer} baseline by 10.5\%. \textit{Coverage-based Grey-box fuzzer} ($k=2, \beta=3$) also outperformed the baseline, yielding a 6.9\% improvement. In contrast, \textit{Coverage-based Black-box fuzzer} ($k=2, \beta=5$) performed slighly worse than the baseline, highlighting a suboptimal choice of the \(\beta\) parameter for this specific target. Similar behavior for certain configurations was also observed in CovFuzz paper. In terms of vulnerability discovery, all selected fuzzers performed comparably, consistently detecting \textbf{C1}, \textbf{C3}, and \textbf{C4} across all runs.

In summary, the evaluation results provide clear answers to \textbf{RQ3} and \textbf{RQ4}. To answer \textbf{RQ3}, the fuzzers adapted from CovFuzz—namely \textit{Random}, \textit{Coverage-based Grey-box}, and \textit{Coverage-based Black-box}—proved effective when applied to the Thread protocol, indicating that the CovFuzz approach generalizes well to this setting. Consistent with the original CovFuzz study, the \textit{Coverage-based Grey-box} fuzzer achieved the highest code coverage and the most reliable vulnerability discovery on both OT-FTD and OT-MTD targets.

With respect to \textbf{RQ4}, the newly introduced \textit{TLV Inserter} fuzzer also showed strong performance. Across all tested configurations, it outperformed both the \textit{Random} baseline and the \textit{Coverage-based Black-box} fuzzer, establishing itself as the most effective \textit{black-box} approach evaluated in this study. These results indicate that \textit{TLV Inserter} is particularly well suited for testing commercial Thread devices, where coverage feedback is unavailable.

\begin{table*}[]
\centering
\caption{Average number of iterations required by each fuzzer to reproduce the crashes.}
\label{tab:CrashPerformance}
\begin{tabular}{|c|c|c|c|c|c|c|c|}
\hline
\textbf{Target} & \textbf{Fuzzer}                  & \textbf{\begin{tabular}[c]{@{}c@{}} Avg. iter.\\ length (s)\end{tabular}} & \textbf{C1}    & \textbf{C2}      & \textbf{C3}       & \textbf{C4}       & \textbf{C5}       \\ \hline
\multirow{12}{*}{\textbf{OT-FTD}} 
& \textbf{Random ($\bm{k=2}$)}      & 0.34                                                                    & 229±238        & 126±91          & \textbf{NO} & \(\bm{-}\)        & 350±300          \\ \cline{2-8}
& \textbf{GB ($\bm{k=2, \beta=1}$)}  & 0.34                                                                    & 56±47          & 200±247    & \textbf{NO}     & \(\bm{-}\)        & 337±991          \\ \cline{2-8}
& \textbf{GB ($\bm{k=2, \beta=3}$)}  & 0.35                                                                    & 137±126        & 148±31          & 3031±462** & \(\bm{-}\)       & 440±306          \\ \cline{2-8}
& \textbf{GB ($\bm{k=2, \beta=5}$)}  & 0.31                                                                    & 197±132        & 168±181         & 3942±994**** & \(\bm{-}\)      & 1410±2948        \\ \cline{2-8}
& \textbf{GB ($\bm{k=2, \beta=7}$)}  & 0.29                                                                    & 138±124        & 193±141****  & \textbf{NO}   & \(\bm{-}\)        & 3290±4325***     \\ \cline{2-8}
& \textbf{BB ($\bm{k=2, \beta=1}$)}  & 0.54                                                                    & 72±94          & 150±162     & \textbf{NO}    & \(\bm{-}\)        & 992±1150         \\ \cline{2-8}
& \textbf{BB ($\bm{k=2, \beta=3}$)}  & 0.62                                                                    & 373±385        & 231±205         & 286±0*     & \(\bm{-}\)        & 360±598          \\ \cline{2-8}
& \textbf{BB ($\bm{k=2, \beta=5}$)}  & 0.55                                                                    & 171±180        & 435±329         & 6122±285**  & \(\bm{-}\)       & 196±153          \\ \cline{2-8}
& \textbf{BB ($\bm{k=2, \beta=7}$)}  & 0.55                                                                    & 76±58          & 580±662   & \textbf{NO}      & \(\bm{-}\)        & 535±384          \\ \cline{2-8}
& \textbf{TLV ($\bm{\gamma = 0.0}$)} & 0.34                                                                    & 644±424        & 215±291         & 4900±338*** & \(\bm{-}\)       & 553±538          \\ \cline{2-8}
& \textbf{TLV ($\bm{\gamma = 0.5}$)} & 0.35                                                                    & 234±140        & 329±269         & 4543±3302** & \(\bm{-}\)       & 434±89           \\ \cline{2-8}
& \textbf{TLV ($\bm{\gamma = 1.0}$)} & 0.36                                                                    & 308±169        & 328±197         & 2006±0*  & \(\bm{-}\)         & 178±131          \\ \hhline{|=|=|=|=|=|=|=|=|}
\multirow{4}{*}{\textbf{OT-MTD}} & \textbf{Random ($\bm{k=2}$)} & 0.68 & 251±132            & \(\bm{-}\)                                                       & 8±5         & 18±15 & \(\bm{-}\)               \\ \cline{2-8}
& \textbf{GB ($\bm{k=2, \beta=3}$)}  & 0.6 & 228±190  & \(\bm{-}\)                                                                & 1±1       & 100±80  & \(\bm{-}\)  \\ \cline{2-8}
& \textbf{BB ($\bm{k=2, \beta=5}$)}  & 0.99 &  428±432 & \(\bm{-}\)                                                                &   20±11    & 52±47 & \(\bm{-}\)  \\ \cline{2-8}
& \textbf{TLV ($\bm{\gamma = 1.0}$)} &                                       0.81                              &    241±159     & \(\bm{-}\)         &  6±7 &   157±78      & \(\bm{-}\)          \\ \hline
\multicolumn{8}{l}{ } \\
\multicolumn{8}{l}{ \hspace{1.5cm} The notation $\bm{\overbrace{*\dots*}^{N}}$ means the crash occurred in exactly \(\bm{N}\) runs (instead of every run).} \\
\multicolumn{8}{l}{\hspace{2.5cm} The notation \textbf{NO} denotes that the crash was not observed in any of the runs.} \\
\multicolumn{8}{l}{\hspace{2.5cm} The symbol \(\bm{-}\) indicates that the target is not vulnerable to the corresponding crash.} \\
\multicolumn{8}{l}{\hspace{1.5cm} \textbf{GB} = \textit{Coverage-based Grey-box}; \textbf{BB} = \textit{Coverage-based Black-box}; \textbf{TLV} = \textit{TLV Inserter}.} \\
\end{tabular}
\end{table*}

\subsection{Comparing ThreadFuzzer with an AFL++-based oracle} \label{subsec:ComparisonWithOSSFuzz}
Since ThreadFuzzer is the first framework capable of fuzzing Thread targets, no prior work exists for direct comparison. However, because OpenThread is actively tested using the OSS-Fuzz framework~\cite{OSS_Fuzz}, it is reasonable to compare ThreadFuzzer against the fuzzing setup currently employed in OpenThread.

For context, OSS-Fuzz aims to improve the security and stability of widely used open-source projects by combining state-of-the-art fuzzing techniques with scalable, distributed execution. OSS-Fuzz supports several modern fuzzing engines, including LibFuzzer~\cite{libfuzzer} and AFL++~\cite{aflplusplus}. To integrate a project into OSS-Fuzz, developers need to provide a \textit{fuzzing harness}—a lightweight wrapper that initializes the target environment and feeds fuzz-generated inputs into a specific function or module under test, thereby enabling effective automated vulnerability discovery.

\subsubsection{Comparing with an OpenThread harness}
As part of the OSS-Fuzz integration, OpenThread provides multiple fuzzing harnesses to test different components of its codebase. Of particular relevance is the harness designed to fuzz incoming radio packets~\cite{otFuzzingHarness}. This harness initializes an OpenThread instance, configures it as a network Leader, and then injects fuzzed input directly into its radio receive path. The pseudocode for this harness is provided in Algorithm~\ref{alg:otFuzzingHarness}.

Before running the fuzzing experiments, the OpenThread OSS-Fuzz configuration disables run-time encryption and integrity checks, effectively creating a white-box fuzzing environment. While this instrumentation enables efficient fuzzing, it requires intrusive modifications to the target code—an approach that is often impractical, sometimes impossible, and frequently undesirable when dealing with commercial or closed-source implementations. In contrast, ThreadFuzzer does not require any modifications to the target code and can operate in a fully \textit{black-box} mode. Consequently, the evaluation described in this section is \textit{not} a standard comparison aimed at demonstrating an incremental improvement over previous work. Rather, it serves to benchmark ThreadFuzzer against a near-ideal baseline.

However, after the initial experiments, it was discovered that the harness implemented by OpenThread is inherently limited. It operates in a \textit{stateless} manner and allows only the initial state of the target to be fuzzed, making the fuzzer incapable of reaching deeper protocol states. As a result, \textit{none} of the vulnerabilities identified by ThreadFuzzer were detected using AFL++ with the harness implemented by OpenThread.

\begin{algorithm}[!htb]
    \caption{Pseudocode of OpenThread fuzzing harness.}\label{alg:otFuzzingHarness}
    \begin{algorithmic}[1]
        \State \textbf{Input:} $fuzzedInput$
        \Comment{Input from a fuzzing engine}
        \State \LeftComment[0\dimexpr\algorithmicindent]{Create and initialize fresh OpenThread instance}
        \State dut = \Call{createOtInstance}{$\ $}
        \LeftComment[0\dimexpr\algorithmicindent]{Promote instance to a network leader}
        \State \Call{otBecomeLeader}{dut}
        \LeftComment[0\dimexpr\algorithmicindent]{Advance virtual time to finalize leader election}
        \State \Call{advanceTime}{$dut, 10$s}
        \LeftComment[0\dimexpr\algorithmicindent]{Feed fuzzed packet into radio receive handler}
        \State \Call{otRadioReceive}{dut, $fuzzedInput$}
        \LeftComment[0\dimexpr\algorithmicindent]{Advance virtual time to observe the after-effects}
        \State \Call{advanceTime}{$dut, 10$s}
    \end{algorithmic}
\end{algorithm}

\begin{algorithm}[htb!]
  \caption{Enhanced OpenThread harness pseudocode.}\label{alg:improvedOtFuzzingHarness1}
  \begin{algorithmic}[1]
    \Procedure{sendFuzzFrameToDUT}{$dut, frame$}
      \LeftComment[0\dimexpr\algorithmicindent]{Wrap raw payload into an MLE frame}
      \State $frame \gets$ \Call{wrapWithMLEHeaders}{$frame$}
      \LeftComment[0\dimexpr\algorithmicindent]{Feed it into radio receive handler}
      \State \Call{otRadioReceive}{$dut, frame$}
    \EndProcedure

    \Statex
    \State \textbf{Input:} $fuzzedInput$ \Comment{Byte array from the fuzzer}
    \Procedure{fuzzingIteration}{$fuzzedInput$}
      \LeftComment[0\dimexpr\algorithmicindent]{Parse control bytes and payload}
      \State $dut\_type \gets fuzzedInput[0]$
      \State $state\_code \gets fuzzedInput[1]$
      \State $payload \gets fuzzedInput[2:]$

      \LeftComment[0\dimexpr\algorithmicindent]{Create DUT instance for the selected type}
      \State $dut \gets$ \Call{createOtInstance}{$dut\_type$}

      \LeftComment[0\dimexpr\algorithmicindent]{Initiate an \textit{Attach procedure}}
      \State \Call{otBecomeChild}{$dut$}

      \State \Call{advanceTime}{$dut, 1$s} \Comment{Wait \textit{ParentRequest}}
      \LeftComment[0\dimexpr\algorithmicindent]{If \texttt{DetachedAfterParentRequest}}
      \If{$state\_code = 0$}
        \State \Call{sendFrameToDUT}{$dut, payload$}
      \EndIf

      \State \Call{otRadioReceive}{$dut, ParentResponse$}
      
      \State \Call{advanceTime}{$dut, 1$s} \Comment{Wait \textit{ChildIDRequest}}
      \LeftComment[0\dimexpr\algorithmicindent]{If \texttt{DetachedAfterChildIDRequest}}
      \If{$state\_code = 1$}
        \State \Call{sendFuzzFrameToDUT}{$dut, payload$}
      \EndIf

      \State \Call{otRadioReceive}{$dut, ChildIDResponse$}
      \If{\Call{otDeviceRole}{$dut$} $\neq$ CHILD}
        \State \Return \Comment{Abort if \textit{DUT} is not a child}
      \EndIf
      \LeftComment[0\dimexpr\algorithmicindent]{If \texttt{Child}}
      \If{$state\_code = 2$}
        \State \Call{sendFuzzFrameToDUT}{$dut, payload$}
      \EndIf
      \LeftComment[0\dimexpr\algorithmicindent]{If \texttt{ChildAfterChildUpdateRequest}}
      \If{$state\_code = 3$}
        \LeftComment[0\dimexpr\algorithmicindent]{Wait \textit{ChildUpdateRequest}}
        \State \Call{advanceTime}{$dut, 235$s}
        \State \Call{sendFrameToDUT}{$dut, payload$}
      \EndIf

      \State \Call{otBecomeRouter}{$dut$}
      
      \State \Call{advanceTime}{$dut, 1$s} \Comment{Wait \textit{AddrSolicit}}
      \LeftComment[0\dimexpr\algorithmicindent]{If \texttt{ChildAfterAddressSolicit}}
      \If{$state\_code = 4$}
        \State \Call{sendFuzzFrameToDUT}{$dut, payload$}
      \EndIf

      \State \Call{sendFrameToDUT}{$dut, AddrSolicitResponse$}
      \If{\Call{otDeviceRole}{$dut$} $\neq$ ROUTER}
        \State \Return \Comment{Abort if DUT is not a Router}
      \EndIf

      \LeftComment[0\dimexpr\algorithmicindent]{If \texttt{Router}}
      \If{$state\_code = 5$}
        \State \Call{sendFuzzFrameToDUT}{$dut, payload$}
      \EndIf

      \LeftComment[0\dimexpr\algorithmicindent]{Wait and observe the final behavior}
      \State \Call{advanceTime}{$dut, 10$s}
    \EndProcedure
  \end{algorithmic}
\end{algorithm}

\subsubsection{Extending an OpenThread harness}
Rather than simply concluding that ThreadFuzzer outperforms the existing OpenThread fuzzing setup, the original harness was extended to support stateful fuzzing. This modification enabled a more meaningful comparison and directly addressed \textbf{RQ5}. As previously noted, AFL++ executed with this extended harness was treated as a state-of-the-art oracle—that is, it was assumed to discover all existing bugs, provided it could be driven into the correct protocol state. The objective was therefore to assess whether ThreadFuzzer could achieve strong performance even under this highly demanding comparison scenario.


Two enhanced variants of the stateful harness were developed: one in which \textit{DUT} acts as a device attempting to join a Thread network, and another where \textit{DUT} serves as a network leader receiving packets from a joining device. Since both variants follow similar design principles, only the former is described in detail here. The corresponding pseudocode is shown in Algorithm~\ref{alg:improvedOtFuzzingHarness1}.

The core idea is to interpret the first byte of the fuzzer-generated input as an indicator of the target DUT type (OT-MTD or OT-FTD) and the second byte as a selector for the desired internal state of the \textit{DUT}. Before injecting the fuzzed payload, the harness initializes the appropriate DUT and transitions it to the specified state by replaying a sequence of pre-constructed, valid packets, thereby making the harness effectively \textit{stateful}. The valid packets were captured manually in advance for replay. Once the \textit{DUT} reaches the intended state, the fuzzed input is injected, after which normal operation resumes with only benign packets being transmitted. To ensure comparability with ThreadFuzzer, all fuzzed inputs are encapsulated with the appropriate lower-layer headers, restricting fuzzing activity to the MLE layer.

The described setup models six distinct internal states of \textit{DUT}, each corresponding to a specific phase of the MLE \textit{Attach process}. The defined states capture both the current MLE state and the last packet sent by the \textit{DUT}. The states are encoded as follows:

\begin{itemize}
    \item[0:] \texttt{DetachedAfterParentRequest}
    \item[1:] \texttt{DetachedAfterChildIDRequest}
    \item[2:] \texttt{Child}
    \item[3:] \texttt{ChildAfterChildUpdateRequest}
    \item[4:] \texttt{ChildAfterAddressSolicit}
    \item[5:] \texttt{Router}
\end{itemize}




\subsubsection{Comparing with an extended OpenThread harness}
For evaluation, 24-hour experiments were performed using the extended harness in combination with AFL++ as the fuzzing engine. The results confirmed that the enhanced harness effectively improved fuzzing performance and enabled the identification of vulnerabilities in the OpenThread library. As expected, AFL++ successfully reproduced all crashes previously detected by ThreadFuzzer and additionally discovered one new crash, referred to as \textit{Crash~6} (\textbf{C6}), which is analyzed together with the others in Section~\ref{subsec:crashAnalysis}.

From this evaluation, it can be concluded that ThreadFuzzer demonstrated its efficiency by successfully identifying five out of six crashes uncovered by a state-of-the-art white-box fuzzer. This result indicates strong overall effectiveness, particularly considering ThreadFuzzer’s capability to operate in a \textit{black-box} setting and its support for fuzzing physical devices.

\subsection{Vulnerability analysis}\label{subsec:crashAnalysis}
This section provides a detailed analysis of all crashes identified during fuzzing, including the five crashes discovered by ThreadFuzzer and one additional crash found by AFL++ using the enhanced OpenThread harness.

\subsubsection{\textbf{Crash 1}}
It is a \textit{reachable assertion} in \textit{DUT} that is triggered upon receiving a malicious \textit{Child ID Response} message. The vulnerability arises when the \texttt{thread\_nwd.tlv.prefix.length} field is set to \textit{255}. Due to the simplicity of the condition required to trigger this crash, all evaluated fuzzers were able to discover it relatively quickly.

\subsubsection{\textbf{Crash 2}}
It is a \textit{stack buffer overflow} that occurs in \textit{DUT} following the reception of a malformed \textit{Child ID Response}. The vulnerability can be triggered by setting the \textit{Length} field in the \textit{Server TLV} to 1, and the \texttt{Address16} field to a value equal to \texttt{Server16}. \textbf{C2} was also successfully found by all the fuzzers, even though on average it took a bit longer to find it. The crash does not happen in OT-MTD as it does not store the \textit{Server TLV}.
It is important to note that this is a \textit{non-crashing} buffer overflow, which does not cause a runtime crash but can be detected using memory sanitization.

\subsubsection{\textbf{Crash 3}}\label{subsubsec:C3}
It is a \textit{reachable assertion} in \textit{DUT} that can be triggered by two different packet types. The vulnerability occurs during the parsing of the \textit{Network Data TLV}, which is present in both \texttt{Data Response} and \texttt{Child ID Response} messages. The \textit{reachable assertion} is triggered when the \texttt{thread\_nwd.tlv.len} field is set to 255, causing the parser to exceed expected bounds when removing data regarding the leader. For FTDs, the crash is less likely since the node can act as a Leader and lack the relevant data, avoiding the vulnerable path. More extensive TLV parsing also increases the chances of dropping malformed packets early.

\subsubsection{\textbf{Crash 4}}
It is a \textit{reachable assertion} in \textit{DUT} that is triggered upon reception of a malicious \textit{Child Update Response} packet, in which the \texttt{mle.tlv.timeout} field is set to an excessively large value (e.g., $4294967295$). The vulnerable code responsible for this behavior is present only in the OT-MTD build. All evaluated fuzzers were able to identify this vulnerability relatively quickly.

\subsubsection{\textbf{Crash 5}}
It is a \textit{reachable assertion} in \textit{DUT} that is triggered under specific timing conditions during the Router promotion process in OT-FTD. To reproduce this crash, a malicious \textit{Advertisement} message must be sent after the device issues an \textit{Address Solicit Request} but before it receives the corresponding \textit{Address Solicit Response}. The injected \textit{Advertisement} manipulates the \textit{Leader Data} by setting the \textit{Leader ID} to 255. When the legitimate \textit{Address Solicit Response} subsequently arrives, the assertion fails due to the inability to locate a valid leader. As OT-MTD devices are not capable of becoming Routers, they are unaffected by \textit{Crash~5}.

\subsubsection{\textbf{Crash 6}}
It is a \textit{stack buffer overflow} in \textit{DUT} that is triggered by a specially crafted \textit{Child ID Response} message in which the \texttt{thread\_nwd.tlv.prefix.length} field is set to 255, as in \textbf{C1}. In addition, the packet must include one optional TLV not present in the default \texttt{Child ID Response} format. While this structure can theoretically be produced by ThreadFuzzer, particularly using \textit{TLV Inserter}, the vulnerability was not discovered during evaluation due to a more subtle requirement.
Specifically, the content of the prefix field must be extended to match the modified \texttt{thread\_nwd.tlv.prefix.length} value—that is, it must contain exactly 256 bits (32 bytes) of data. This level of precise content shaping is beyond the current capabilities of the fuzzers implemented in ThreadFuzzer. In theory, \textit{TLV Inserter} could satisfy this constraint by inserting a TLV of the appropriate size directly after the prefix field, such that the inserted TLV is interpreted as part of the prefix. However, satisfying all required conditions—modifying the prefix length, aligning the actual prefix content accordingly, and appending a valid optional TLV—requires a rare and highly specific combination of mutations. As a result, the likelihood of this crash being triggered by ThreadFuzzer with the currently implemented fuzzing algorithms is extremely low.

\begin{table*}[htb!]
\centering
\caption{List of tested commercial devices}
\label{tab:commercialDUTresults}
\begin{tabular}{|c|c|c|c||c|c|c|c|c|c|}
\hline
\textbf{DUT}   &
\textbf{\makecell{Software\\version}} & \textbf{Type} & \textbf{OT} &
\textbf{C1}  & \textbf{C2} & \textbf{C3}  & \textbf{C4} & \textbf{C5} & \textbf{C6} \\
\hline        
\makecell{\textbf{Eve Systems Door and Window Matter}} & \makecell{3.2.1}
& MTD & YES & YES & \(\bm{-}\) & YES & NO & \(\bm{-}\) & NO* \\
\hline
\makecell{\textbf{Aqara Door and Window Sensor P2}} & \makecell{1.0.1.1}
& MTD & YES & YES & \(\bm{-}\) & YES & NO & \(\bm{-}\) & NO* \\
\hline
\makecell{\textbf{Arre Contact Sensor}} & \makecell{1.0.0+0}
& MTD & NO & NO & \(\bm{-}\) & NO & NO & \(\bm{-}\) & NO* \\
\hline
\makecell{\textbf{Nanoleaf Essentials Matter Smart Bulb}} & \makecell{3.2.0}
& FTD & YES & YES & NO* & NO & \(\bm{-}\) & YES & NO*\\
\hline
\multicolumn{10}{c}{} \\
\multicolumn{10}{c}{The notation NO* indicates a non-crashing buffer overflow that does not produce a detectable runtime failure.} \\
\multicolumn{10}{c}{The symbol \(\bm{-}\) indicates that the target type is not vulnerable to the corresponding crash.}
\\
\end{tabular}
\end{table*}

\section{Fuzzing the commercial devices}\label{subsec:CommercialFuzz}
ThreadFuzzer is the only available framework capable of fuzzing physical Thread devices. This section therefore evaluates its effectiveness on commercial Thread devices.
The analysis consider four commercial Thread products from different vendors, comprising three MTDs and one FTD. Although all but one these devices publicly indicate the use of OpenThread~\cite{OpenThread} in their products, the framework itself is agnostic to the underlying implementation. ThreadFuzzer can thus be applied to any Thread-compliant device, irrespective of whether OpenThread is used internally.

This section is divided into two parts. The first examines whether the vulnerabilities uncovered by ThreadFuzzer during the simulation-based fuzzing (C1–C5) and one extra revealed by AFL++ with the extended harness (C6) can be reproduced on commercial Thread devices. The second presents proof-of-concept fuzzing campaigns on selected devices and compares their outcomes with the simulation-based results.

\subsection{Reproducing the crashes C1-C6}


To assess whether crashes C1–C6 are reproducible on commercial devices, the packet mutations that triggered these crashes in the simulation-based experiments are applied during the experiments with commercial devices.
The results are summarized in Table~\ref{tab:commercialDUTresults}. For each device, the table reports the name, software version, device type (MTD or FTD), whether it is based on OpenThread, and the observed crash behavior.

The experiments yield several key observations. First, the device not based on OpenThread did not exhibit any of the crashes. Second, \textbf{C2} and \textbf{C6}, both stack buffer overflows, were not observed on any device due to their non-crashing nature. Third, \textbf{C1} and \textbf{C5}—both reachable assertion failures—were reproducible across all OpenThread-based devices. Additionally, all OpenThread MTDs consistently crashed under \textbf{C3}, whereas the tested FTD did not—likely due to the lower likelihood of this crash occurring in FTDs, as explained in Section~\ref{subsubsec:C3}. Finally, \textbf{C4} could not be reproduced on any of the devices; without access to internal details, it is unclear whether the vulnerability is mitigated or never exposed under real-world configurations.

In response to \textbf{RQ2}, these results suggest that most crashing vulnerabilities (i.e., reachable assertions) found in the OpenThread library are reproducible on commercial devices that rely on OpenThread. In contrast, non-crashing memory safety issues are more difficult—if not impossible—to detect with the \textit{black-box} fuzzing setups. This finding reinforces the challenges of fuzzing embedded physical devices and aligns with prior observations in the domain~\cite{What_You_Corrupt_Is_Not_What_You_Crash}.

\subsection{Performance and efficiency of commercial device fuzzing}
To illustrate the performance characteristics of fuzzing physical devices, two representative devices—one OpenThread-based and one not—were selected. It is argued that evaluating a single OpenThread-based device is sufficient to obtain a representative view of fuzzing performance on physical devices that rely on OpenThread.

Both devices were subjected to three fuzzing campaigns of \(500\) iterations each. Given the similar vulnerability-discovery performance observed across different fuzzers for MTD targets, the \textit{Random} fuzzer with \(k = 2\) was selected as the fuzzing engine.
The epoch size (as defined in Algorithm~\ref{alg:realDeviceFuzzingProcedure}) was empirically set to \(N = 4\), which provides a good balance between performance and stability. This means that after every four fuzzing iterations, a clean MLE attach procedure is executed to obtain the reboot-count values, followed by a factory reset of the device before starting the next epoch.

Table~\ref{tab:commercialDUTfuzzing} reports the average iteration duration and the average number of iterations required to discover the relevant crashes, following the format of Table~\ref{tab:CrashPerformance}.

\begin{table*}[htb!]
\centering
\caption{Performance comparison of ThreadFuzzer on two commercial devices, one based on OpenThread and one using a proprietary implementation.}    
\label{tab:commercialDUTfuzzing}
\begin{tabular}{|c|c|c|c|c|c|c|c|c|c|}
\hline
\textbf{DUT}  & \textbf{Type} & \textbf{OT}                & \textbf{\begin{tabular}[c]{@{}c@{}} Avg. iter.\\ length (s)\end{tabular}}  & 
\textbf{C1}  & \textbf{C2} & \textbf{C3}  & \textbf{C4} & \textbf{C5} & \textbf{C6} \\
\hline        
\makecell{\textbf{Eve Systems Door and Window Matter}}  & MTD & YES & 45 & 78±75 & \(\bm{-}\) & 13±10 & \(\bm{-}\) & \(\bm{-}\) & \(\bm{-}\) \\
\hline
\makecell{\textbf{Arre Contact Sensor}} & MTD & NO  
& 45 & \(\bm{-}\) & \(\bm{-}\) & \(\bm{-}\) & \(\bm{-}\) & \(\bm{-}\) & \(\bm{-}\) \\
\hline
\multicolumn{9}{c}{} \\
\multicolumn{9}{c}{The symbol \(\bm{-}\) indicates that the crash did never occur.}
\\
\end{tabular}
\end{table*}

According to the analysis in the previous subsection, an OpenThread-based MTD is expected to exhibit \textbf{C1} and \textbf{C3}, which is consistent with the outcomes of these campaigns. The data further aligns with simulation-based experimental results from Table~\ref{tab:CrashPerformance}, as \textbf{C3} is typically reached more quickly than \textbf{C1}. For the non-OpenThread device, the absence of crashes (\textbf{C1}–\textbf{C6}) matches the earlier observations, and no additional crashes were identified during the fuzzing campaign.

In terms of runtime, fuzzing a commercial device requires roughly 45 seconds per iteration on average. This is approximately 66 times slower than simulation-based fuzzing under identical parameters. This slowdown stems from the factors discussed in Section~\ref{subsec:DUT}, namely the use of virtual time in the simulator, which speeds up the fuzzing, and the need for stabilization delays and a full, clean reboot of the physical device before retrieving reboot-count information, which depending on the device can be more or less time-consuming.

\section{Discussion}
\subsection{Impact of the Discovered Vulnerabilities}
To exploit the vulnerabilities presented in this paper, a malicious device must already be a part of the Thread network—that is, it must have successfully completed the commissioning process, which is outside the scope of this work. The vulnerabilities caused by \textit{reachable assertions} result in temporary denial-of-service (DoS) conditions on the victim device. However, as long as the attacker remains in the network, the attack can be repeated indefinitely, leading to a persistent DoS. The severity of such an attack depends on the criticality of the targeted device and the context in which it operates. The impact of the stack buffer overflow vulnerabilities depends largely on whether they can be escalated to remote code execution (RCE). To the best of current knowledge, none of the stack buffer overflows identified in this study are exploitable for RCE, which significantly limits their impact.

\subsection{Limitations and Future Work}
This section discusses the main limitations of the current study and outlines potential directions for future work.

\subsubsection{Fuzzing Restricted to the MLE Layer}
ThreadFuzzer currently supports packet interception and manipulation exclusively at the MLE layer, which is arguably the most distinctive layer introduced in the Thread protocol stack. Consequently, only the MLE layer implementation of \textit{DUT} is subject to fuzzing. However, the techniques presented in this work are generalizable and can be extended to enable fuzzing of other layers within the Thread protocol stack.

\subsubsection{Lack of Correlated Field Detection}
ThreadFuzzer does not incorporate any mechanism for detecting correlations between protocol fields. For example, the field \texttt{thread\_nwd.tlv.prefix.length} is semantically correlated with \texttt{thread\_nwd.tlv.prefix}, as the size of the prefix must match the specified length. The absence of such dependency awareness limits the ability to generate semantically valid but nontrivial inputs, and is the primary reason why ThreadFuzzer failed to uncover \textit{Crash~6}.

\subsubsection{Manipulation Always Starting from a Benign Packet}\label{subsec:benign_packet_limitation}
ThreadFuzzer is designed to perform fuzzing by always mutating benign, well-formed packets. This design imposes inherent limitations on the structure of the resulting fuzzed packets, which cannot deviate drastically from the original intercepted packet. Although \textit{TLV Inserter} introduced in this work partially mitigates this constraint by allowing insertion of additional TLVs, the fundamental limitation remains: entirely synthetic or structurally unrelated packets cannot be generated under the current model. This limitation is another reason why \textbf{C6} was not discovered by ThreadFuzzer.

\subsubsection{No Mechanism to Avoid Repeated Crashing Mutations}
ThreadFuzzer does not track mutations that lead to crashes during execution. As a result, the same crashing input—produced by repeatedly mutating a benign packet in a similar way—may occur numerous times during the fuzzing campaign. This may repeatedly trigger the same vulnerability, hindering code path exploration and reducing fuzzing efficiency. This issue is particularly evident when fuzzing OT-MTD, where \textbf{C3} occurs frequently, causing repeated \textit{DUT} restarts and longer iteration times.

\subsubsection{Physical Device Support Restricted to Matter-Compliant Devices}
As described in Section~\ref{subsec:DUT}, when fuzzing physical devices, ThreadFuzzer relies on the cluster from  Matter protocol to detect crashes. As a result, the current implementation is limited to devices that support Matter. For context, according to OpenThread~\cite{OpenThread}, 66 out of 100 products that use OpenThread also support Matter.

\section{Conclusion}
This paper introduces \textit{ThreadFuzzer}—the first fuzzing framework specifically designed for systematically testing implementations of the Thread protocol. It supports both virtual OpenThread nodes and commercial Thread devices by intercepting and manipulating packets at the MLE layer.

The framework integrates multiple fuzzing strategies, adopting the \textit{Random} and \textit{Coverage-based} fuzzers from CovFuzz, and introducing a new \textit{TLV Inserter} tailored to the TLV-based structure of MLE messages. These fuzzers are thoroughly evaluated on the OpenThread library, resulting in the discovery of five previously unknown vulnerabilities. Their relative efficiency was also assessed. The results show that the fuzzers proposed in CovFuzz generally transfer well to the Thread protocol. However, the TLV inserter introduced in this work provides a noteworthy addition, emerging as the best-performing \textit{black-box} fuzzing engine within the ThreadFuzzer framework.

To assess the practical performance, ThreadFuzzer is benchmarked against a white-box state-of-the art fuzzer AFL++ using an extended version of the OSS-Fuzz OpenThread harness. It successfully reproduces five out of six AFL++-discovered crashes, demonstrating its overall effectiveness while highlighting key limitations that inform future development of network protocol fuzzers.







\clearpage
\bibliographystyle{ieeetr}
\bibliography{references}

\appendices

\section{Data Availability}
To ensure transparency and reproducibility of the results presented in this work, the fuzzing framework together with the presented data is publicly available at~\url{https://anonymous.4open.science/r/ThreadFuzzer-BC12}. Researchers and practitioners are encouraged to explore and utilize it in compliance with the terms of the repository’s license.

\section{Formulas for the mutation probability changes from CovFuzz}~\label{app:Formulas}
This section details the formulas used by the \textit{Coverage-based fuzzer}s incorporated into ThreadFuzzer from CovFuzz. The descriptions are adapted from the original CovFuzz work.

Equation~\ref{eq:prob_change} describes the change of the mutation probability for the field $f$ that was mutated during the $i$-th fuzzing iteration.
\begin{equation}\label{eq:prob_change}
    p^{i}_f \leftarrow p^{i-1}_f + \frac{G(c^{(i)},i)}{\log_2(|V_f| + 1)}
\end{equation}

The function \(G(c^{(i)}, i)\)—scaled by the logarithm of the total number of possible values for the field \(f\)—plays a central role in Equation~\ref{eq:prob_change}. It depends on the iteration $i$ and the new coverage $c^{(i)}$ found during iteration $i$.

The formula for $G(c^{(i)},i)$ can be seen in Equation~\ref{eq:F}, combined with Equations~\ref{eq:f_1} and~\ref{eq:g}:
\begin{equation}\label{eq:F}
    G(c^{(i)},i) = \frac{ h(c^{(i)}) \cdot g(i)}{n^{(i)}}
\end{equation}
\begin{equation}\label{eq:f_1}
    h(c^{(i)}) =
    \begin{cases}
        1, & \text{if}\ c^{(i)} > 0 \\
        -1, & \text{otherwise}
    \end{cases}
\end{equation}
\begin{equation}\label{eq:g}
\begin{split}
    g(i) = 
    \begin{cases}
    \beta \cdot \gamma(i), & \text{if}\ c^{(i)} > 0 \\
    \frac{1}{\beta \cdot \gamma(i)}, & \text{otherwise}
    \end{cases}
    \\
    \text{where}\ 
    \gamma(i) = \min(\frac{i}{warm\_i}, 1)
\end{split}
\end{equation}
where $n^{(i)}$ represents the total number of mutated fields during iteration $i$ and functions $h(c^{(i)})$ and $g(i)$ depend on the new coverage found and the fuzzing iteration number, respectively.
Normalization of the probability adjustment is achieved by dividing by $n^{(i)}$—the number of mutated fields.

In Equation~\ref{eq:g}, $warm\_i$ and $\beta$ are user-defined hyper-parameters. The parameter $warm\_i$ represents the \textit{warming} phase of \textit{Coverage-based fuzzer}, defining the number of iterations considered as the stage of fuzzing, where finding new coverage is relatively easy. During this \textit{warming} phase, $\gamma(i)$ remains less than 1, resulting in a more gradual adjustment of probabilities when new coverage is found and a more aggressive adjustment when no new coverage is discovered. Once $i$ exceeds $warm\_i$, meaning the \textit{warming} phase has ended, $\gamma(i)$ is always set to $1$ and $g(i)$ no longer depends on the iteration count.
According to a suggestion from CovFuzz, ThreadFuzzer uses \(warm_i = 2000\).

The parameter $\beta$ controls the rate at which mutation probabilities are adjusted. Higher values of $\beta$ result in larger increments to the probabilities of mutated fields when new coverage is achieved, while reducing the decrease when coverage remains unchanged. Conversely, lower values of $\beta$ lead to smaller adjustments to mutation probabilities upon new coverage, with larger reductions when coverage stagnates. Following the recommendations from CovFuzz, multiple experiments were conducted to determine an appropriate value for the parameter \(\beta\) in Section~\ref{subsec:ThreadEval}.



\end{document}